\documentclass[journal,onecolumn,comsoc]{IEEEtran}
\IEEEoverridecommandlockouts
\usepackage[T1]{fontenc}
\usepackage[dvips]{psfrag}
\usepackage[dvips]{epsfig}
\usepackage{epic,color}
\usepackage[cmex10]{amsmath}
\usepackage{amsfonts}
\usepackage{amssymb}
\usepackage{amsmath}
\usepackage[cmintegrals]{newtxmath}
\usepackage{bm}
\usepackage{yfonts}
\usepackage{array}
\usepackage{bigstrut}
\usepackage{multirow}
\usepackage{dsfont}
\usepackage{cite}
\usepackage{fixltx2e}
\usepackage{accents}
\usepackage{mathtools}
\usepackage[normalem]{ulem}
\usepackage{enumerate}
\usepackage{stfloats}
\usepackage[T1]{fontenc}
\usepackage{float}
\usepackage{epstopdf}
\usepackage{mathtools}
\usepackage{algorithm}
\usepackage{algorithmicx}
\usepackage{algpseudocode}
\usepackage{algorithm,algpseudocode,float}
\usepackage{lipsum}
\usepackage{soul,xcolor}
\usepackage{xparse}
\usepackage{acronym}
\usepackage{soul}
\usepackage{graphicx}
\NewDocumentCommand{\ceil}{s O{} m}{%
  \IfBooleanTF{#1} 
    {\left\lceil#3\right\rceil} 
    {#2\lceil#3#2\rceil} 
}
\NewDocumentCommand{\floor}{s O{} m}{%
  \IfBooleanTF{#1} 
    {\left\lfloor#3\right\rfloor} 
    {#2\lfloor#3#2\rfloor} 
}

\newcommand{\mbf}[1]{\mathbf{#1}}

\algrenewcommand\algorithmicrequire{\textbf{Input:}}
\algrenewcommand\algorithmicensure{\textbf{Output:}}

\usepackage{graphicx}
\usepackage{subfigure}
	
\IEEEpubid{\begin{minipage}{\textwidth}\ \\[12pt]
  1558-2558  \copyright \, 2020 IEEE. Personal use is permitted, but republication/redistribution requires IEEE permission.
See https://www.ieee.org/publications/rights/index.html for more information.
\end{minipage}} 

\begin{document}

\title{A Novel Spectrally-Efficient Uplink Hybrid-Domain NOMA System}
\author{Chen Quan, Animesh Yadav, \IEEEmembership{Senior Member, IEEE}, Baocheng Geng, Pramod K. Varshney, \IEEEmembership{Life Fellow, IEEE} and H. Vincent~Poor, \IEEEmembership{Life Fellow,~IEEE}
\thanks{This work was supported in part by the U.S. National Science Foundation under Grants CCF-
0939370 and CCF-1908308.

C. Quan, A. Yadav, B. Geng and P. K. Varshney are with Department of Electrical Engineering and Computer Science, Syracuse University, Syracuse, NY, 13244, USA. (e-mail: \{chquan, ayadav04, bageng, varshney\}@syr.edu).

H. Vincent Poor is with  the Department of Electrical Engineering, Princeton University, Princeton, NJ 08544, USA. (e-mail: poor@princeton.edu)}}

\maketitle
\IEEEpubidadjcol
\begin{abstract}
This paper proposes a novel hybrid-domain (HD) non-orthogonal multiple access (NOMA) approach to support a larger number of uplink users than the recently proposed code-domain NOMA approach, i.e., sparse code multiple access (SCMA). HD-NOMA combines the code-domain and power-domain NOMA schemes by clustering the users in small path loss (strong) and large path loss (weak) groups. The two groups are decoded using successive interference cancellation while within the group users are decoded using the  message passing algorithm. To further improve the performance of the system, a spectral-efficiency maximization problem is formulated under a user quality-of-service constraint, which dynamically assigns power and subcarriers to the users. The problem is non-convex and has sparsity constraints. The alternating optimization procedure is used to solve it iteratively. We apply successive convex approximation and reweighted $\ell_1$ minimization approaches to deal with the non-convexity and sparsity constraints, respectively. The performance of the proposed HD-NOMA is evaluated and compared with the conventional SCMA scheme through numerical simulation. The results show the potential of HD-NOMA in increasing the number of uplink users that can be supported in a given time-frequency resource.
\end{abstract}


\IEEEpeerreviewmaketitle

\vspace{-0.5cm}

\section{Introduction}

\IEEEPARstart{I}{n} the midst of spectrum scarcity, the fifth generation (5G) and beyond networks are expected to support a higher demand for wireless connections, which could be realized by one of the potential bandwidth efficient technologies known as non-orthogonal multiple access (NOMA) \cite{fang2016lattice,yadav2019IEEEWCM}. The two common types of NOMA are power-domain NOMA (PD-NOMA)\cite{saito2013vtc,islam2016IEEECST}, and code-domain NOMA (CD-NOMA) \cite{taherzadeh2014scma}\cite{nikopour2013sparse}, which enables multiple users to successfully transmit data on the same time-frequency resource, and hence, the improved spectral-efficiency (SE). 

Although conventional PD-NOMA and CD-NOMA schemes provide higher SE, efforts have been made to further improve the SE. Towards this end, in \cite{moltafet2017new}, a novel power domain SCMA (PSMA) scheme based on the conventional SCMA is proposed to enhance the SE with a reasonable increase in complexity. In \cite{li2016joint,evangelista2019IEEETWC}, a joint codebook assignment and power allocation design for uplink SCMA is discussed  for improving the SE. In \cite{sharma2019joint}, a combination of PD-NOMA and SCMA is proposed for improving the downlink channel capacity by grouping users according to their channel gains. 

However, the aforementioned works have not investigated the optimal design of power and subcarrier allocation while increasing the number of users over the uplink channel. In this paper, we propose a scheme to support a large number of users over the uplink channel. The main contributions of this paper are as follows:
\begin{itemize}
    \item A novel HD-NOMA scheme  is proposed that combines the PD-NOMA and SCMA schemes to accommodate more users over the uplink channel.
    \item For improved performance, a sum rate optimization problem is formulated that optimally assigns power and subcarriers to the uplink users.
    \item Iterative successive convex approximation (SCA) and reweighted $\ell_1$ minimization based approaches to handle non-convex and sparsity constraints, respectively, are used to solve the formulated non-convex optimization problem for rapid convergence.
\end{itemize}

The paper is organized as follows. Section I presents the system model of the proposed scheme. Section III presents the problem formulation. Section IV discusses the proposed solution and the algorithm. Section V presents and discusses numerical simulation results and Section VI concludes the paper.

\section{SYSTEM MODEL} Consider an  uplink system where a set $\mathcal{J}=\{j; j=1,\dots, J\}$ of single antenna users communicate synchronously and simultaneously with a single antenna base station (BS) over a set $\mathcal{K}=\{k; k=1,\dots, K\}$ of orthogonal subcarriers using the NOMA scheme. In the sequel, the channel coefficient corresponding to the channel between user $j$ and the BS on subcarrier $k$ is denoted by $h_{j,k}\, \forall j\in \mathcal{J}, \, \forall k\in \mathcal{K}$, which includes the path loss and Rayleigh fading effects  and assumed to be block-stationary and perfectly known at the BS. For the sake of convenience, we briefly discuss the conventional PD-NOMA and SCMA systems before the proposed HD-NOMA system.
\subsection{Uplink PD-NOMA System}
In uplink PD-NOMA, users with different channel gains superimpose their data and transmit simultaneously on the same subcarrier. Different channel gains essentially help in decoding the users by employing the SIC technique. In this technique, from the received signal, the user with the highest channel gain is decoded first and then its reconstructed signal is subtracted from the received signal. Then the next user with the next highest channel gain is decoded and so on until the last user \cite{tse2005fundamentals}. 
\subsection{Uplink SCMA System}
\begin{figure}[ht]
\vspace{-0.2in}
\centering
\subfigure[$J = 6$, $K = 4$, $d_{v} = 3$, $d_{f} = 2$ and $\lambda = 1.5$.]{%
\includegraphics[height=5.5em]{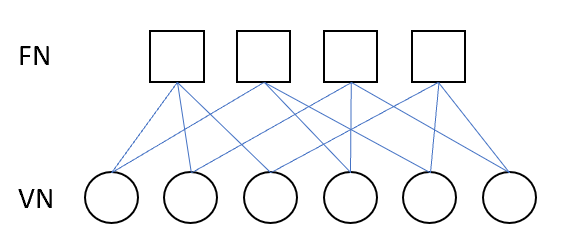}
\label{fig:FactorGraph}}
\,
\subfigure[$J_{\text{s}} = J_{\text{w}} = 6$, $K = 4$, $d_{v}^{\text{s}} = d_{v}^{\text{w}} = 3$, $d_{f}^{\text{s}} = d_{f}^{\text{w}} = 2$ and $\lambda = 3$.]{%
\includegraphics[height=8em]{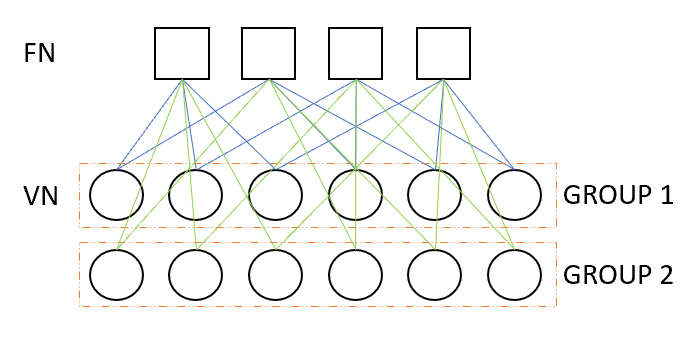}
\label{fig:FactorGraph_HD-NOMA}}

\caption{Factor graphs of (a) conventional SCMA and (b) HD-NOMA.}
\label{fig:figure}
\end{figure}

 In uplink SCMA, each user transmits over $d_f$ out of $K$ available subcarriers, and hence, the SCMA encoder maps the $\log_{2}{M}$ bits to the $d_f$-sparse\footnote{A codeword with $d_f$ non-zero elements.} $K$-dimensional codewords which come from a predefined user-specific codebook of size $M$, e.g., the $j^{th}$ user codebook is denoted as $B_j=\{\mathbf{x}_{j,1}, \ldots,\mathbf{x}_{j,m}, \ldots, \mathbf{x}_{j,M}\}$ where $\mathbf{x}_{j,m} \in \mathbb{C}^{K\times 1}$. At most $d_{v}= \tbinom{K-1}{d_f-1}$ users can transmit simultaneously on the same subcarrier, and the maximum number of users supported by the system is $J = \tbinom{K}{d_{f}}$, which ensures the uniqueness of each user's codebook. The user-to-subcarrier mapping can be easily represented by a factor graph representation. The complete mapping information can be compactly expressed into a factor graph matrix $\mathbf{F}_{J\times K}$ and each of its element $f_{j,k}\in\{0,1\}$ indicates whether (if 1) or not (if 0) the user $j$ uses a subcarrier $k$. An illustrative factor graph is shown in Fig. \ref{fig:FactorGraph}, where $\lambda = J/K$ denotes the overloading factor of the system, and $\text{FN}$ and $\text{VN}$ denote the function and the variable nodes to represent subcarriers and users, respectively. For decoding the received signal, the BS uses a message passing algorithm (MPA) \cite{xiao2015simplified} \cite{kschischang2001factor}. 
 
\vspace{-0.15in}     
\subsection{Uplink HD-NOMA System Model}
In the proposed HD-NOMA system, we take advantage of both PD- and CD-NOMA by combining them to achieve the goal of increasing the system capacity.
\begin{figure}[htbp]
\vspace{-0.15in}
\centerline{\includegraphics[height= 10em]{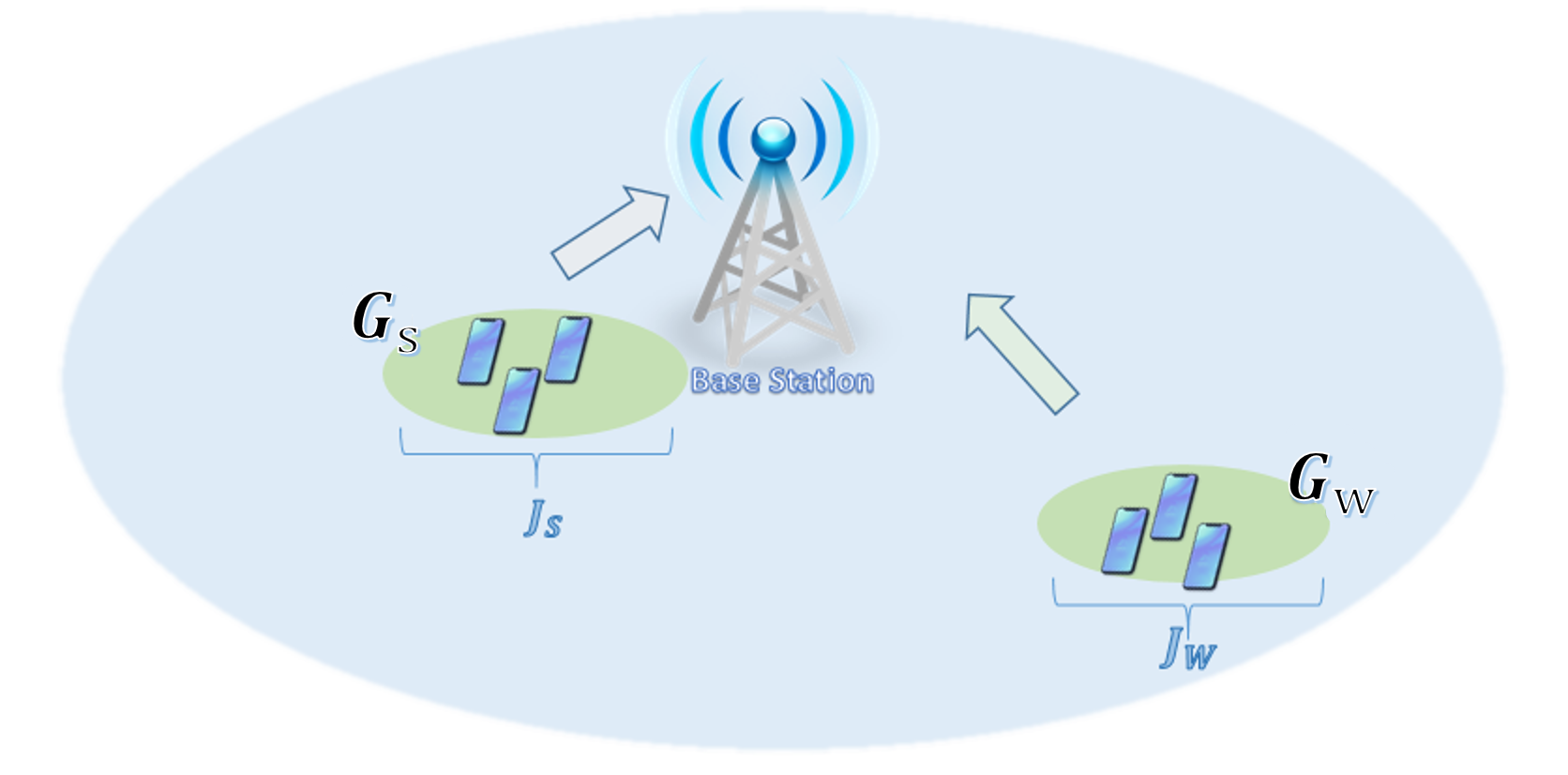}}
\vspace{-0.1in}
\caption {System model of the proposed scheme  with $(J_{\text{s}} + J_{\text{w}})$ users.}
\label{fig:SystemModel_HD-NOMA}
\end{figure}

We assume that the number of users in the set $\mathcal{J}$ for HD-NOMA is higher than the one assumed in PD- and CD-NOMA systems, and divided into two subsets, i.e., $\mathcal{J}_{\text{s}}=\{j;j=1,\ldots,J_{\textbf{s}}\}$ and $\mathcal{J}_{\text{w}}=\{j;j=1,\ldots,J_{\textbf{s}}\}$. The sets  $\mathcal{J}_{\text{s}}$ and $\mathcal{J}_{\text{w}}$ consist of users that are experiencing small (strong user group $G_{\text{s}}$) and large (strong user group $G_{\text{w}}$) path losses, respectively, as shown in Fig, \ref{fig:SystemModel_HD-NOMA}. The groups $G_s$ and $G_w$ use $d_{f}^{\text{s}}$-sparse and $d_{f}^{\text{w}}$-sparse sets of codebooks $\mathcal{B}_{\text{s}}$ and $\mathcal{B}_{\text{w}}$, respectively. Like the conventional SCMA system, there are at most $J_{\text{s}}=\tbinom{K}{d_{f}^{\text{s}}}$
and $J_{\text{w}}=\tbinom{K}{d_{f}^{\text{w}}}$ users in the strong and weak groups, respectively. $J_s$ and $J_w$ are not necessarily equal here. Fig. \ref{fig:FactorGraph_HD-NOMA} shows one possible factor graph of HD-NOMA. 

For this system, the received signal $\bold{y}\in \mathbb{C}^{K\times 1}$ can be expressed as 
\begin{equation}
    \bold{y} = \sum_{i=1}^{J_{\text{s}}} \sqrt{p^{\text{s}}}\text{diag}(\bold{h}_{i}^{\text{s}})\bold{x}_{i}^{\text{s}} + \sum_{j=1}^{J_{\text{w}}} \sqrt{p^{\text{w}}}\text{diag}(\bold{h}_{j}^{\text{w}})\bold{x}_{j}^{\text{w}}+ \bold{n},
\end{equation}
where $\mathbf{x}_i^{\text{s}}\in {B}_i^{\text{s}},\, \forall i \in \mathcal{J}_{\text{s}}$ and $\mathbf{x}_j^{\text{w}}\in {B}_j^{\text{w}}, \, \forall j \in \mathcal{J}_{\text{w}}$ are the codewords corresponding to the strong and weak users, respectively. $\mathbf{h}_i^{\text{s}}=[h_{i,1}^{\text{s}}, \ldots, h_{i,K}^{\text{s}}]^T$ and $\mathbf{h}_j^{\text{w}}=[h_{j,1}^{\text{w}}, \ldots, h_{j,K}^{\text{w}}]^T$ denote the channel vectors of $i^{th}$ strong and  $j^{th}$ weak users, respectively. Since the users in each group will experience approximately the same path loss, we assume that the users in the same group use the same power to transmit their signals. Therefore, $p^{\text{s}}$ and $p^{\text{w}}$, are the transmit powers and $d_{f}^{\text{s}}$ and $d_{f}^{\text{w}}$ are the number of subcarriers allocated to the strong and weak users, respectively. $d_{v}^{\text{s}}$ and $d_{v}^{\text{w}}$ are the numbers of strong and weak users in each subcarrier, respectively. $\bold{n} \sim \mathcal{CN}(\mathbf{0}, \sigma_n^2 \bold{I})$ is the independent and identically distributed (iid) additive white Gaussian noise (AWGN) vector.

\begin{figure}[htbp]
 \vspace{-0.2in}
\centerline{\includegraphics[height=6em]{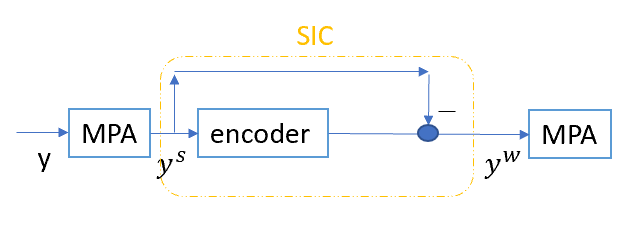}}
\vspace{-0.1in}
\caption {Block diagram for decoding for HD-NOMA.}
\label{fig:Decoding_HD-NOMA}
\end{figure}

Note that, decoding all the users at once leads to exponentially increased computational complexity with the number of users in the system. However, the channel gain based user clustering allows us to employ the SIC method and a sequential MPA that has lower computational complexity compared with the conventional SCMA system with the same number of users. For example, the computational complexity of the SCMA is $\mathcal{O}(M^{d_v^{\text{w}}+d_v^{\text{s}}})$, while the one for our proposed HD-NOMA system is $\mathcal{O}(\max\{M^{d_v^{\text{w}}}, M^{d_v^{\text{s}}}\})$. The decoding procedure uses the MPA and SIC decoding as depicted in Fig. \ref{fig:Decoding_HD-NOMA}.

\paragraph{Decoding of strong users}
Given the channel state information $\bold{h}_{i}^{\text{s}}=[h^{\text{s}}_{i,1}, \ldots, h^{\text{s}}_{i,K}]\, \forall i\in \mathcal{J_{\textbf{s}}}$ and  $\bold{h}_{j}^{\text{w}}=[h^{\text{w}}_{j,1}, \ldots, h^{\text{w}}_{j,K}]\, \forall j\in \mathcal{J_{\textbf{w}}}$, and with the assumption that $h_i^{\text{s}}\sqrt{p^{\text{s}}}>> h_j^{\text{w}}\sqrt{p^{\text{w}}}$, the codewords corresponding to the strong users are decoded first by employing the MPA. The data of weak users is regarded as Gaussian noise with variance $p^{\text{w}} |h_j^{\text{w}}|^2$. Afterwards, SIC is performed to prepare the estimated signal $\bold{y}_{\text{sic}}^{\text{w}}\in \mathbb{C}^{K\times 1}$ for decoding of weak users as $\bold{y}_{\text{sic}}^{\text{w}} = \bold{y} - \sum_{i=1}^{J_{\text{s}}} \text{diag}(\bold{h}_{i}^{\text{s}})\bold{x}_{i}^{\text{s}}\sqrt{p^{\text{s}}}.$

\paragraph{Decoding of weak users}
The codewords for weak users are decoded by employing the MPA on the signal $\bold{y}_{\text{sic}}^{\text{w}}$. 

\section{PROBLEM FORMULATION}
By utilizing channel diversity to further enhance the performance of the system, we optimize the power allocation to the two groups and subcarrier assignment within groups. From Fig. \ref{fig:Decoding_HD-NOMA}, we note that the decoding procedure consists of two sequential steps. So the decoding performance of the second MPA is guaranteed on the premise that the performance of the first MPA is guaranteed. Therefore, the objective is to maximize the sum rate for the strong group subject to meeting the quality-of-service (QoS) requirement of the weak group. The idea is to decrease the interference introduced by the weak group in the first step while also ensures an improved decoding performance in the second step. 

The general achievable sum rate of $J$-user SCMA system is derived in \cite{evangelista2019IEEETWC} as:
\begin{equation}\label{eq:sumrate_scma}
\text{SR} = \sum_{k=1}^K \log_{2}\Big( 1 + \frac{\sum_{j=1}^{J} p |h_{j,k}|^2f_{j,k}}{\sigma_n^2}\Big),
\end{equation}
where each user transmits with power $p$.  $f_{j,k}$ represents the mapping between user $j$ and subcarrier $k$, i.e., $f_{j,k}=1$ means user $j$ is mapped to subcarrier k, otherwise it is 0. Now using \eqref{eq:sumrate_scma} and owing to the decoding procedure shown in Fig. \ref{fig:Decoding_HD-NOMA}, the achievable sum rate of both $J_{\text{s}}$-user strong and $J_{\text{w}}$-user weak groups can be expressed, respectively, as:
\begin{equation}
    \text{SR}^{\text{s}}(\bold {p,F}) = \sum_{k=1}^K\log_2\Bigg(1+\frac{\sum_{i=1}^{J_{\text{s}}}|h^{\text{s}}_{i,k}|^2f^{\text{s}}_{i,k}p^{\text{s}}}{\sum_{j=1}^{J_{\text{w}}}|h^{\text{w}}_{j,k}|^2f^{\text{w}}_{j,k}p^{\text{w}}+\sigma_n^2}\Bigg),
\end{equation}
\begin{equation}
    \text{SR}^{\text{w}}(\bold {p,F}) =\sum\limits_{k=1}^K\log_2\Bigg(1+\frac{\sum_{j=1}^{J_{\text{w}}}|h^{\text{w}}_{j,k}|^2f^{\text{w}}_{j,k}p^{\text{w}}}{\sigma_n^2}\Bigg),
\end{equation}
where $\bold p$ includes $\{p^{\text{s}},p^{\text{w}}\}$, and $\bold F$ includes $\{\bold F^{\text{s}}_{J_{\text{s}}\times K},\bold F^{\text{w}}_{J_{\text{w}}\times K}\}$ the factor graph matrices for the strong and weak groups, respectively. Note that the factor graph matrix is essentially a subcarrier assignment indication matrix, where $\bold F^{\text{s}}=[[f^{\text{s}}_{1,1},\ldots, f^{\text{s}}_{J_{\text{s}},1}]^T\cdots\allowbreak [f^{\text{s}}_{1,K},\ldots, f^{\text{s}}_{J_{\text{s}},K}]^T]$ and a similar matrix for the weak user group.

Now, the optimization problem is written as follows:

\begin{subequations}\label{original_eq}
\begin{align}
\underset{\bold {p,F}}{\text{max}} \quad & \text{SR}^{\text{s}}(\bold {p,F})\label{eq:objectiveF}\\
\text{s.t.} \quad & \text{SR}^{\text{w}}(\bold {p,F}) \geq \bar{R}^{\text{w}}\label{eq:WU_sumrate}\\
 & \sum\limits_{k=1}^{K}f^{\text{s}}_{i,k} \leq d_f^{\text{s}}\, \forall	i\in \mathcal{J}_{\text{s}},\, \sum\limits_{k=1}^{K}f^{\text{w}}_{j,k} \leq d_f^{\text{w}}\, \forall	j\in \mathcal{J}_{\text{w}} \label{eq:subcarrier}\\
 &\sum\limits_{i=1}^{J_{\text{s}}}f^{\text{s}}_{i,k} \leq d_v^{\text{s}}\, \forall k\in \mathcal{K},\, \sum\limits_{j=1}^{J_{\text{w}}}f^{\text{w}}_{j,k} \leq d_v^{\text{w}}\, \forall k\in \mathcal{K} \label{eq:Users}\\
 &\sum_{k=1}^{K}f^{\text{s}}_{i,k}p^{\text{s}}\leq P, \, \sum\limits_{k=1}^{K}f^{\text{w}}_{i,k}p^{\text{w}}\leq P\,\, \forall i\in \mathcal{J}_{\text{s}} \cup \mathcal{J}_{\text{w}},\label{eq:power}\\
 & f^{\text{s}}_{i,k}, f^{\text{w}}_{j,k}\in \left\{0,1\right\} \, \forall i\in \mathcal{J}_{\text{s}},\forall j\in \mathcal{J}_{\text{w}}, \forall k \in \mathcal{K}\label{eq:binary}
\end{align}
\end{subequations}
where, $P$ is the original maximum transmitting power for each user.

The constraint \eqref{eq:WU_sumrate} ensures that the sum rate of the weak group is greater than a certain minimum threshold $\bar{R}^{\text{w}}$ so that the signal from the weak users could be successfully decoded. Constraints in \eqref{eq:subcarrier} restrict the number of sub-carriers used per user, and constraints in \eqref{eq:Users} restrict the number of users per sub-carrier. \eqref{eq:power} restricts the maximum power allowed to be used by each user. However, coupled variables and the non-convexity of problem (\ref{original_eq}) make it mathematically intractable to solve.

\section{PROPOSED SOLUTION}
Due to the presence of non-convexity and sparse binary matrix variables in \eqref{eq:objectiveF}, \eqref{eq:WU_sumrate}, \eqref{eq:power}, and \eqref{eq:binary}, respectively, (\ref{original_eq}) is difficult to solve optimally with fast convergence in a reasonable time. Thus, we obtain the sub-optimal solutions by developing a rapidly converging algorithm. Further, since the variables $\mathbf{p}$ and $\mathbf{F}$ have non-overlapping domains, we solve (\ref{original_eq}) via the alternating optimization (AO) method, which consists of fixing one set of variables and then optimizing the other, iteratively. If each subproblem is convex, then the AO algorithm is able to converge to the optimal solution \cite{bertsekas1997nonlinear}. Thus, we relax the original problem and the details are shown in the next subsections.

\subsection{Optimization of $\mathbf{p}$}
After introducing a slack variable $t_1$ and fixing $\mathbf{F}$, \eqref{original_eq} can be written as the following subproblem:
\begin{align}\label{eq:DC_p_slack}
\underset{\mathbf{p},t_1}{\max} \{t_1|\text{SR}^{\text{s}}(\bold {p}) \geq t_1, \eqref{eq:WU_sumrate},\, \text{and}\, \eqref{eq:power}\}.
\end{align}
Note that the constraint $\text{SR}^{\text{s}}(\bold {p}) \geq t_1$ in \eqref{eq:DC_p_slack} is non-convex as it can be written in a difference-of-convex (d.c.) functions form as $u_1(\bold p)-v_1(\bold p)\geq t_1$, where $u_1(\bold p)= \sum_{k=1}^K\log_2\Big(\sum_{i=1}^{J_{\text{s}}}|h^{\text{s}}_{i,k}|^2f^{\text{s}}_{i,k}p^{\text{s}}+\sum_{j=1}^{J_{\text{w}}}|h^{\text{w}}_{j,k}|^2f^{\text{w}}_{j,k}p^{\text{w}}+\sigma_n^2\Big)$ and $v_1(\bold p)= \sum_{k=1}^K\log_2\Big(\sum_{j=1}^{J_{\text{w}}}|h^{\text{w}}_{j,k}|^2f^{\text{w}}_{j,k}p^{\text{w}}+\sigma_n^2\Big)$ are both concave functions. To make this constraint convex, we linearize $v(\mathbf{p})$ using first-order Taylor approximation around the point of operation, $\mathbf{p}^{(n_1)}$, as follows:
\begin{equation}\label{eq:DC_constraints_p}
   u_1(\bold p)-v_1(\bold p^{(n_1)})-\langle\nabla v_1({\bold p^{(n_1)}}),(\bold p-\bold p^{(n_1)})\rangle\geq t_1,
\end{equation}
where $\nabla v_1({\bold p})=\Big[\frac{\partial v_1({\bold p})}{\partial p^{\text{s}}}, \frac{\partial v_1({\bold p})}{\partial p^{\text{w}}}\Big]$ and $\langle\cdot,\cdot\rangle$ represents the inner product operation. Now, with this relaxed constraint, \eqref{eq:DC_p_slack} is transformed into a convex problem as: 
\begin{align}\label{eq:DC_p_slack_approx}
\underset{\bold {p,t_1}}{\max}\{t_1|\eqref{eq:DC_constraints_p}, \eqref{eq:WU_sumrate},\, \text{and}\, \eqref{eq:power}\},
\end{align}
which can be solved iteratively via the SCA method where the iteration index is represented by $n_1$.

\subsection{Optimization of $\bold {F}$}
Next, after introducing another slack variable $t_2$ and fixing $\mathbf{p}$, \eqref{original_eq} can be written as the following subproblem: 
\begin{align}
\underset{\bold {F,t_2}}{\max}\{t_2|\text{SR}^{\text{s}}(\mathbf{F}) \geq t_2, \eqref{eq:WU_sumrate}-\eqref{eq:binary}\}.
\end{align}

In order to relax sparse binary matrix constraints \eqref{eq:binary}, we utilize the \emph{reweighted} $\ell_1$-\emph{minimization} method \cite{candes2008enhancing} and formulate a locally concave penalty function that more closely resembles the $\ell_0$ norm, which iteratively pushes the matrix $\bold F$ to be sparse. Note that the constraints \eqref{eq:subcarrier} and \eqref{eq:Users} have been changed to  \eqref{eq:subcarrier_f} and \eqref{eq:Users_f} to prevent $\bold F$ from becoming too sparse to satisfy the requirements of \eqref{original_eq}. The relaxed problem can be written as:
\begin{subequations}\label{eq:original_f}
\begin{align}
\hspace{-0.1cm} \underset{\mathbf{F},t_2}{\max}\quad & t_2 - \lambda \sum_{k=1}^K\Big(\sum_{i=1}^{J_{\text{s}}} w^{\text{s},(n_2)}_{i,k}f^{\text{s}}_{i,k}+\sum_{j=1}^{J_{\text{w}}} w^{\text{w},(n_2)}_{j,k}f^{\text{w}}_{j,k}\Big)\label{eq:subproblem2}\\
 \text{s.t.}\quad & \text{SR}^{\text{s}}(\mathbf{F}) \geq t_2, \label{eq:WU_sumrate_f}\\
& \sum\limits_{k=1}^{K}f^{\text{s}}_{i,k} \geq d_f^{\text{s}}\, \forall	i\in \mathcal{J}_{\text{s}},\, \sum\limits_{k=1}^{K}f^{\text{w}}_{j,k} \geq d_f^{\text{w}}\, \forall	j\in \mathcal{J}_{\text{w}} \label{eq:subcarrier_f}\\
&\sum\limits_{i=1}^{J_{\text{s}}}f^{\text{s}}_{i,k} \geq d_v^{\text{s}}\, \forall k\in \mathcal{K},\, \sum\limits_{j=1}^{J_{\text{w}}}f^{\text{w}}_{j,k} \geq d_v^{\text{w}}\, \forall k\in \mathcal{K},\label{eq:Users_f} \\
& 0 \leq f^{\text{s}}_{i,k},f^{\text{w}}_{j,k} \leq 1 \quad \forall i\in \mathcal{J}_{\text{s}},\, \forall j \in \mathcal{J}_{\text{w}}, \forall k \in \mathcal{K},\label{eq:binary_relax}\\
& \eqref{eq:WU_sumrate}\ \text{and}\, \eqref{eq:power}, \label{eq:rest_eqns}
\end{align}
\end{subequations}
where $w^{\text{s},(n_2)}_{i,k}=1/(|f^{\text{s},(n_2-1)}_{i,k}|+\epsilon), \forall	i\in \mathcal{J}_{\text{s}}, \forall k \in \mathcal{K}$ and $w^{\text{w},(n_2)}_{j,k}=1/(|f^{\text{w},(n_2-1)}_{j,k}|+\epsilon), \forall	j\in \mathcal{J}_{\text{w}}, \forall k \in \mathcal{K}$ . $\epsilon$ is a small value. Constant $\lambda$ is the weight of the penalty term.
Due to the constraint \eqref{eq:WU_sumrate_f}, the problem \eqref{eq:original_f} is non-convex. Hence, we linearize it using first-order Taylor approximation as described previously. After linearization,
\eqref{eq:original_f} is transformed into a convex problem as:
\begin{subequations}\label{eq:DC_f_slack_approx}
\begin{align}
\underset{\bold {F},t_2}{\max} \quad & \eqref{eq:subproblem2}\\
\text{s.t.} \quad & u_2(\bold F)-v_2(\bold F^{(n_2)})-\text{tr}\Big(\nabla v_2({\bold F}^{(n_2)})(\bold F-\bold F^{(n_2)})\Big)\geq t_2\\
&\eqref{eq:subcarrier_f}-\eqref{eq:rest_eqns},
\end{align}
\end{subequations}
where 
$u_2(\bold F)= \sum_{k=1}^K\log_2\Big(\sum_{i=1}^{J_{\text{s}}}|h^{\text{s}}_{i,k}|^2f^{\text{s}}_{i,k}p^{\text{s}}+\sum_{j=1}^{J_{\text{w}}}|h^{\text{w}}_{j,k}|^2f^{\text{w}}_{j,k}p^{\text{w}}+\sigma_n^2\Big)$, $v_2(\bold F)= \sum_{k=1}^K\log_2\Big(\sum_{j=1}^{J_{\text{w}}}|h^{\text{w}}_{j,k}|^2f^{\text{w}}_{j,k}p^{\text{w}}+\sigma_n^2\Big)$ are both concave functions, and $\left\{\nabla v_2({\bold F})\right\}_{k,i}=\frac{\partial v_2({\bold F})}{\partial f_{k,i}}$ for $ \forall i\in \mathcal{J}_1 \cup \mathcal{J}_2, \forall k \in \mathcal{K}$. \eqref{eq:DC_f_slack_approx} is also solved iteratively via the SCA method with iteration index represented by $n_2$. 
The pseudocode for solving \eqref{original_eq} approximately is outlined in Algorithm 1. Since the objective function is upper bounded due to power constraints, the proposed AO algorithm generates a monotonic non-decreasing sequence of objective function values, and hence, the sequence converges. Let $n_1^{max}$, $n_2^{max}$ denote the maximum number of iterations for the solution of subproblem \eqref{eq:DC_p_slack_approx} and the solution of subproblem \eqref{eq:DC_f_slack_approx} at the time of convergence for one iteration of the AO algorithm, respectively. The complexity of our proposed algorithm for one iteration of the AO algorithm is $\mathcal{O}\big(\max\{n_1^{max}, n_2^{max}\}(JK+2)^3\big)$.


\renewcommand{\baselinestretch}{1}{
\begin{algorithm}[H]
\caption{The Proposed AO Algorithm}
\label{algo:AO}
\begin{algorithmic}[1]
\Require $\{\mbf{h}^{\text{s}}_i\}_{i=1}^{J_{\text{s}}},\, \{\mbf{h}^{\text{w}}_j\}_{j=1}^{J_{\text{w}}},\, \bar{R}^{\text{w}},\, \epsilon$
\Ensure $\mbf{p}^{*}$, $\mbf{F}^{*}$.
\State Initialize $n:=0$, $\mbf{p}^{(0)}$, $\mbf{F}^{(0)}$;
\Repeat
	\State Set $n_1=0$;
	\Repeat	
		\State \parbox[t]{\dimexpr\linewidth-\algorithmicindent}{Given $\mbf{F}^{(n)}$  optimize (8) for $\mbf{p}^{*}$;}
		\State Update: $\mbf{p}^{(n+1)}=\mbf{p}^{(n_1+1)}=\mbf{p}^{*}$, $n_1:=n_1+1$;
	\Until{Local convergence of $\mbf{p}$}
	\State \parbox[t]{\dimexpr 0.95\linewidth-\algorithmicindent}{Set $n_2 = 0$, $w^{\text{s},(n_2)}_{i,k} = w^{\text{w},(n_2)}_{j,k} = 1,\forall i\in \mathcal{J}_{\text{s}},\,\forall j\in \mathcal{J}_{\text{w}}, \forall k\in \mathcal{K}$;}
	\Repeat 
    \State \parbox[t]{\dimexpr\linewidth-\algorithmicindent}{Given $\mbf{p}^{(n+1)}$ optimize (11) for $\mbf{F}^{*}$;}
    \State \parbox[t]{\dimexpr 0.9\linewidth-\algorithmicindent}{Update: $\mbf{F}^{(n+1)}=\mbf{F}^{(n_2+1)}=\mbf{F}^{*}$, $w^{\text{s},(n_2+1)}_{i,k}=\frac{1}{f^{\text{s},(n_2)}_{i,k}+\epsilon}$, $w^{\text{w},(n+1)}_{j,k}=\frac{1}{f^{\text{w},(n_2)}_{j,k}+\epsilon}$, $\forall i\in \mathcal{J}_{\text{s}},\forall j\in \mathcal{J}_{\text{w}},\forall k\in \mathcal{K}$, and $n_2:=n_2+1$;}
	\Until{Local convergenve of $\mbf{F}$}
	\State Update: $n:=n+1$;
\Until{Convergence of the AO algorithm}
\end{algorithmic}
\end{algorithm}}
\vspace{-0.3cm}
\section{Simulation results}
Numerical simulation results are presented and discussed in this section. The performances in terms of the uncoded bit error rate (BER) and the sum rate are compared between the conventional SCMA , the PD-NOMA and the proposed system. The codebooks are designed according to \cite{taherzadeh2014scma}. The parameter values are set as follows: $\epsilon = 10^{-3}$, $M=4$ and $K = 4$. $J_{\text{w}} = 6$, $J_{\text{s}} = 6$, $d_v^{\text{s}} = d_v^{\text{w}} = 3$, $d_f^{\text{w}} = d_f^{\text{s}} = 2$ for HD-NOMA, and $J = 6$, $d_v = 3$, $d_f = 2$ for the conventional SCMA. The bandwidth $BW$ is 1 MHz. $\sigma_n^2 = -204 + 10\log_{10}(BW/K)$ dB. Path Loss is denoted as $PL = 145.4+37.5\log_{10}(d_i)$, where $d_i$ (in kilometers) is the propagation distance of the signal. We assume $d_i$ to be 0.3 for the strong group and 0.8 for the weak group.
\begin{figure}[t]
\centerline{\includegraphics[width=0.5\linewidth,height=5 cm]{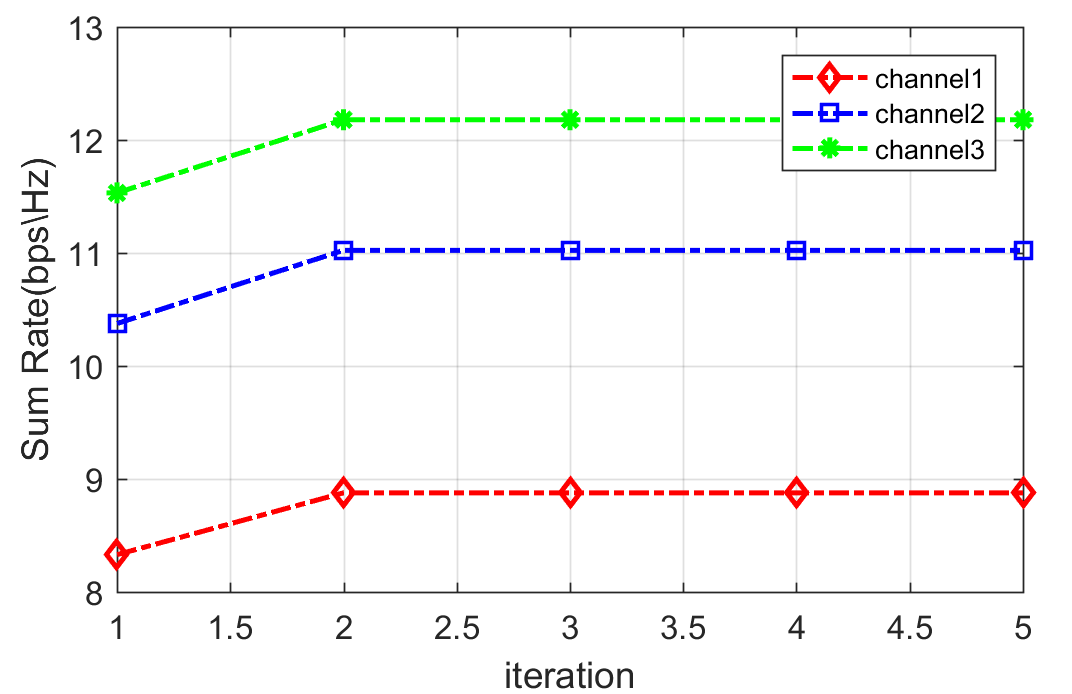}}
\caption{Convergence of the proposed AO algorithm with three different power levels \{30 dBm, 35 dBm and 40 dBm\} and channel realizations.}
\label{fig:convergence}
\end{figure} 
Fig. \ref{fig:convergence} shows the convergence behavior of the iterative Algorithm 1. The objective function values are plotted versus the number of iterations for three independent random channel realizations and power values $P=\{30, 35, 40\}$ dBm. It can be observed that the Algorithm 1 converges in less than three iterations for all three channel realizations. The average run times to converge for the proposed and branch-and-bound (BnB) algorithms are 5 and 500 seconds, respectively, on an x64-based processor with 1.8 GHz clockspeed and 16 Gb RAM.

\begin{figure}[b]
\centerline{\includegraphics[width=0.5\linewidth]{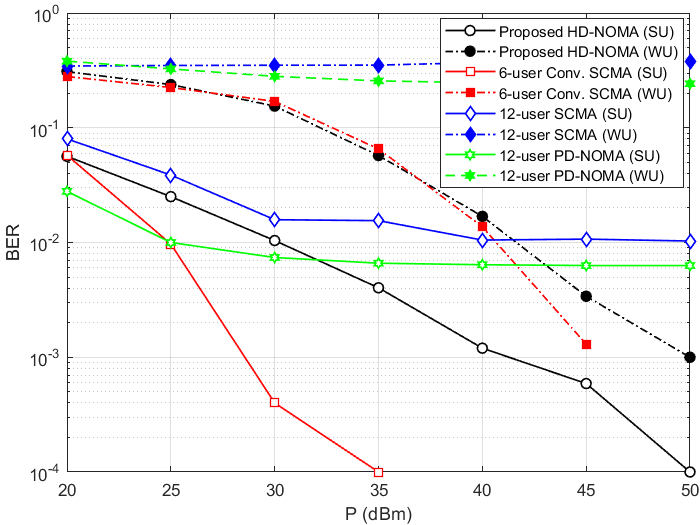}}
\caption{Uncoded BER versus user transmit power.}
\label{fig:BER_performance}
\end{figure}
Fig. \ref{fig:BER_performance} shows the comparison of the uncoded BER performance with varying values of the user transmit power. The BER performance of the proposed HD-NOMA is compared with that of its six-user counterpart in the six-user conventional SCMA, the twelve-user PD-NOMA and the twelve-user SCMA. It can be observed that the strong users in HD-NOMA have lower error rates than their counterparts in the twelve-user conventional SCMA. The weak users in HD-NOMA have lower error rates than their counterparts in the twelve-user conventional SCMA and similar error rates as their counterparts in the six-user conventional SCMA. This is because the users face more interference in the twelve-user conventional SCMA system. However, the reason why the BER performance of the users in HD-NOMA deteriorates compared to their counterparts in the six-user conventional SCMA is that we introduce interferences based on the decoding procedure shown in Fig. \ref{fig:Decoding_HD-NOMA}. Although HD-NOMA results in relatively higher error rates compared to the conventional six-user SCMA model, it is within acceptable rates considering the fact that a higher number of users are supported by HD-NOMA. It can also be observed that the proposed HD-NOMA outperforms the conventional SCMA and PD-NOMA in BER performance with the same number of users.

\begin{figure}[t]
\centerline{\includegraphics[width=0.5\linewidth]{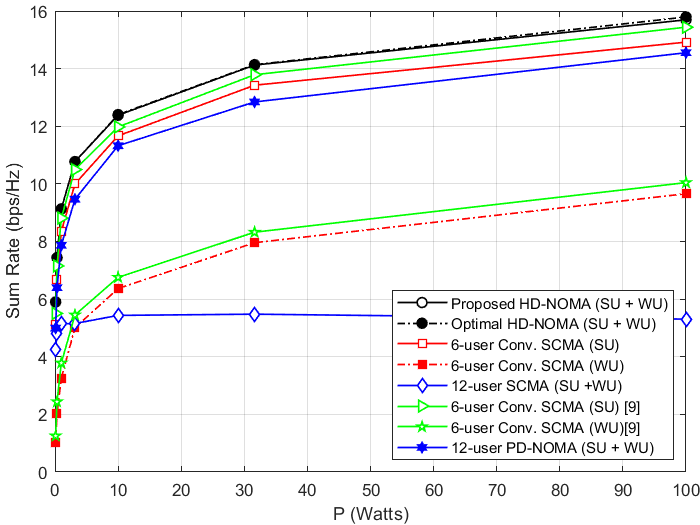}}
\caption{Sum rate versus user transmit power.}
\label{fig:sumrate}
\end{figure}
Fig. \ref{fig:sumrate} compares the sum-rate performance of HD-NOMA with the proposed solution and the optimal solution obtained via BnB algorithm, the six-user conventional SCMA (with equal power and optimized according to \cite{evangelista2019IEEETWC}), the twelve-user PD-NOMA and SCMA systems with varying user transmit power. It can be observed that the HD-NOMA scheme supports much higher data rate and outperforms both SCMA and PD-NOMA (including twelve and six-users) systems. Moreover, the optimality gap between the optimal solution and our proposed solution is negligible.

\section{Conclusion And Discussion}
In this paper, we have proposed a novel HD-NOMA scheme for uplink channels that overcomes the limitation on the number of users previously set in the conventional SCMA scheme. HD-NOMA combines the conventional PD-NOMA and SCMA schemes by clustering the users in strong and weak user groups respectively. The groups are decoded by the use of SIC as they have significant differences in their channel gains while the within the group users are decoded using an MPA. Sum rate performance is improved by jointly designing the power and subcarrier allocation. Numerical results show improved SE over the conventional SCMA scheme and demonstrate its potential for supporting larger number of users in spectrum congested networks.

\bibliographystyle{IEEEtran}

\end{document}